\documentclass{jpsj-suppl}
\usepackage{bm}
\usepackage{epstopdf}
\usepackage{graphicx}
\usepackage{txfonts} 
\title{Microscopic Study of $\alpha+N$ Bremsstrahlung from Effective and Realistic Inter-nucleon Interactions}

\author{J\'er\'emy \textsc{Dohet-Eraly}$^{1,2}$, Sofia \textsc{Quaglioni}$^{3}$, Petr \textsc{Navr\'atil}$^{1}$, and Guillaume \textsc{Hupin}$^{3}$}

\inst{$^1$TRIUMF, 4004 Wesbrook Mall, Vancouver, BC V6T 2A3, Canada\\
$^2$Physique Quantique, C. P. 165/82, and Physique Nucl\'eaire Th\'eorique et Physique Math\'ematique, C.P. 229, Universit\'e libre de Bruxelles (ULB), B-1050 Brussels, Belgium\\
$^3$ Lawrence Livermore National Laboratory, P.O. Box 808, L-414, Livermore, California 94551, USA}

\email{jdoheter@triumf.ca}

\abst{Based on an effective nucleon-nucleon interaction, a microscopic cluster model of the nucleus-nucleus
bremsstrahlung, including implicitly a part of the effects of meson-exchange currents via
an extension of the Siegert theorem is applied to the $\alpha+p$ and $\alpha+n$ systems. The contributions of the
$E1$ and $E2$ transitions to the bremsstrahlung cross sections are evaluated and their relative importance
for the mirror systems $\alpha+p$ and $\alpha+n$ is compared. Another approach based on realistic two- and three-nucleon interactions and the No-Core Shell Model/Resonating-Group
Method is also investigated. Some preliminary results for the $\alpha+p$ bremsstrahlung are displayed.}
\kword{nuclear bremsstrahlung, $\alpha+N$, calculated cross sections, microscopic model}

\begin{document}
\maketitle

\section{Introduction}
Nucleus-nucleus bremsstrahlung is a radiative transition between continuum states where the
photon emission is induced by a nuclear collision. Interest in this process has recently been revived by the experimental study of electromagnetic transitions in the unstable ${^8{\rm Be}}$ via the $\alpha+\alpha$ bremsstrahlung~\cite{DCK13} and by the perspective of using the $t(d,n\gamma)\alpha$ bremsstrahlung to diagnose plasmas in fusion experiments~\cite{MBB01}.

The study of the $\alpha+N$ bremsstrahlung is motivated by several reasons.
First, it makes possible a direct comparison between theory and experiment since the $\alpha+p$ system is one of the few light-ion systems for which bremsstrahlung cross sections were measured~\cite{WHM71}. 
Second, the $\alpha+n$ bremsstrahlung is a necessary preliminary step to the study of the $t(d, n\gamma)\alpha$ bremsstrahlung since it describes the final channel. Finally, the $\alpha+N$ elastic scattering is very well described by the microscopic cluster models and the more complex but more fundamental \textit{ab initio} methods \cite{HLN13}.

The description of the electromagnetic transitions in nuclear systems is based on the interaction between the electromagnetic field of the photon and the nuclear current, which is due to the motion of the nucleons and also to the motion of the mesons, responsible for the nucleon-nucleon (NN) and nucleon-nucleon-nucleon (NNN) interactions.
However, the contribution of the meson-exchange currents was neglected in most previous studies of nucleus-nucleus bremsstrahlung. Recently, it has been proposed~\cite{DEB13} to include partially the meson-exchange currents in the bremsstrahlung models by using an extended version of the Siegert theorem~\cite{SWA90}, which does not rely on the long-wavelength approximation (LWA). 
This approach has been applied as well in microscopic models~\cite{DEB13,DE14} as in potential models~\cite{DEB14}.
It has to be noted that the LWA cannot be made in the continuum-to-continuum transitions because it leads to divergent matrix elements of the electric transition multipole operators and thus, divergent bremsstrahlung cross sections, since the initial and final states are not square-integrable.

In addition to the implicit inclusion of the meson exchange currents, using the extended Siegert
theorem reduces the complexity of the calculations, making easier the development of \textit{ab initio}
bremsstrahlung models.
\section{The $\alpha+N$ bremsstrahlung cross section}
An $\alpha$ particle and a nucleon collide at the initial relative momentum $\bm{p}_i=\hbar\bm{k}_i$ 
in the $z$ direction and relative energy $E_i=p^2_i/2\mu_M$ where $\mu_M$ is the reduced mass of the system. 
After emission of a photon with energy $E_\gamma=\hbar k_\gamma c$, the system has a final relative momentum $\bm{p}_f=\hbar\bm{k}_f$ in the direction $\Omega_f=(\theta_f,\varphi_f)$ 
and a relative energy $E_f=p^2_f/2\mu_M$, which satisfies
\begin{equation}\label{consEn}
E_f = E_i - E_\gamma,
\end{equation}
where the small recoil energy is neglected.
The $\alpha$ particle is assumed to be in its ground state before and after the photon emission. 
Its spin is zero. 
The spin projection of the nucleon before and after the collision, denoted respectively $\nu_i$ and $\nu_f$, can be different.

The bremsstrahlung cross section is evaluated from the multipole matrix elements, which are proportional to the matrix elements of the electromagnetic transition multipole operators $\mathcal{M}^\sigma_{\lambda\mu}$ 
between the incoming initial state $\Psi^{\nu_i +}_i$ in the $z$ direction with energy $E_i$ and the outgoing final state $\Psi^{\nu_f -}_f(\Omega_f)$ with energy $E_f$ and direction $\Omega_f$,
\begin{equation}\label{matu}
u^{\sigma \nu_i \nu_f}_{\lambda\mu}(\Omega_f)=\alpha^\sigma_\lambda \langle\Psi^{\nu_f -}_f(\Omega_f)|\mathcal{M}^\sigma_{\lambda\mu}|\Psi^{\nu_i +}_i\rangle,
\end{equation}
where $\sigma=E$ corresponds to an electric multipole and $\sigma=M$ corresponds to a magnetic multipole and $\alpha^\sigma_\lambda$ is given by 
\begin{equation}
\alpha^E_\lambda=-i \alpha^M_\lambda=-\frac{\sqrt{2\pi(\lambda+1)}i^\lambda k^\lambda_\gamma}{\sqrt{\lambda(2\lambda+1)}(2\lambda-1)!!}.
\end{equation}
Assuming that the photon helicity and the final spin projections are not observed and that the incident beam is unpolarized, the angle-integrated bremsstrahlung cross section is given by~\cite{DE14}
\begin{equation}
\frac{d\sigma}{d E_\gamma } =\frac{E_\gamma p^2_f}{2 \pi^2\hbar^5 c 4\pi\epsilon_0} \sum_{\nu_i \nu_f}\sum_{\sigma\lambda\mu}
\int^\pi_0\frac{|u^{\sigma  \nu_i \nu_f}_{\lambda\mu}(\theta_{f},0)|^2}{2\lambda+1} \sin\theta_f  d\theta_f.
\end{equation}
The explicit form of the electric transition multipole operators $\mathcal{M}^{E}_{\lambda\mu}$ in the Siegert approach for a microscopic model can be found in~\cite{DE14}.
The contribution of the magnetic transitions, which is expected to be weak for the $\alpha+N$ bremsstrahlung at low photon energy, is neglected.
\section{Microscopic approaches}\label{Secmic}
The microscopic description of the $\alpha+N$ system relies on the internal five-body Schr\"odinger equation
\begin{equation}\label{Sch}
H\Psi=E_T\Psi,
\end{equation}
where $H$ is the microscopic internal Hamiltonian, $\Psi$ is the internal wave function, and $E_T$ is the total energy of the system in the center-of-mass (c.m.)\ frame. 
The microscopic internal Hamiltonian $H$ is given by
\begin{equation}\label{defH}
H=\sum^5_{i=1} \frac{p^2_i}{2m_N}+\sum^5_{i>j=1} v_{ij}+\sum^5_{i>j>k=1} v_{ijk}-T_{\rm c.m.},
\end{equation}
where $\bm{p}_i$ is the momentum of nucleon $i$, $m_N$ is the nucleon mass, $v_{ij}$ and $v_{ijk}$ are the two- and three-body potentials describing the NN and NNN interactions between nucleons $i$ and $j$ or $i$, $j$, and $k$, and $T_{\rm c.m.}$ is the c.m.\ kinetic energy. 

The initial and final states $\Psi^{\nu_i +}_i$ and $\Psi^{\nu_f -}_f$ in Eq.~\eqref{matu} are solutions of the Schr\"odinger equation~\eqref{Sch} corresponding to relative energies $E_i$ and $E_f$, respectively, and having the appropriate asymptotic behavior of an incoming or outgoing wave function. 
These states are described following two different approaches: an effective cluster approach, namely the Generator Coordinate Method (GCM)~\cite{Ho77}, and a more realistic cluster approach, namely the No-Core Shell Model/Resonating-Group Method (NCSM/RGM)~\cite{QN09}.
In the GCM, the $\alpha$ cluster wave function is simply the internal wave function of
the $\alpha$ ground state within the harmonic oscillator shell model. In the NCSM/RGM, the $\alpha$ cluster wave
functions are NCSM solutions of the four-nucleon Schr\"odinger equation, where the same inter-nucleon
interaction as in Eq.~\eqref{defH} is considered.
In both approaches, the Microscopic $R$-matrix Method~\cite{BHL77,DB10} is used to enforce the expected asymptotic behavior of the collision wave function.

The inter-nucleon potentials $v_{ij}$ and $v_{ijk}$ must be adapted to the considered approach.
In the GCM approach, an effective NN interaction, the Minnesota potential~\cite{TLT77} complemented by the Coulomb potential, is used. 
No three-body potential is included.
By adjusting the exchange parameter and the spin-orbit strength of the Minnesota potential, the GCM reproduces nicely the experimental elastic phase shifts.
In the NCSM/RGM approach, a version of the NN interaction from the chiral effective field theory at next-to-next-to-next-to-leading order~\cite{EM03} complemented by a local form of the chiral
NNN interaction at next-to-next-to-leading order~\cite{Na07} is first softened by the similarity renormalization
group and then, applied in the calculations.
More details can be found in~\cite{HLN13}.
\section{Results}
\begin{figure}[ht]
\begin{minipage}{18pc}
\includegraphics[width=18pc]{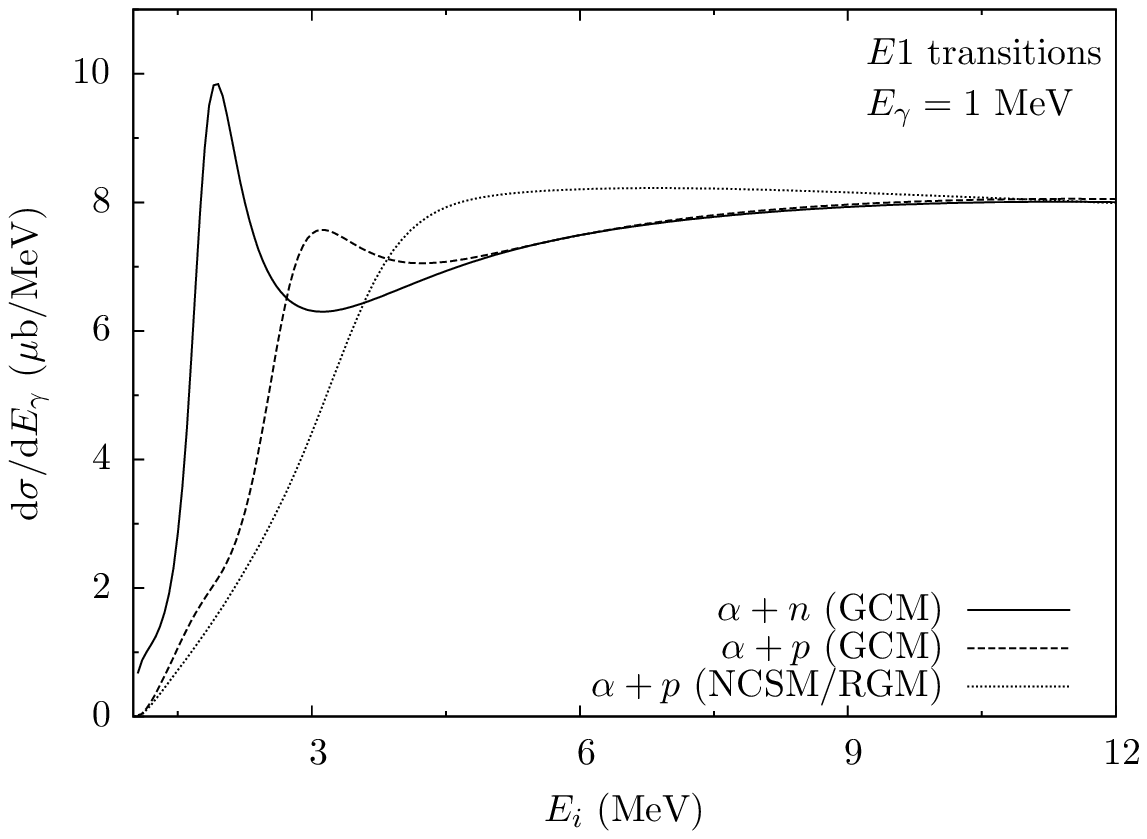}
\caption{\label{E1} The $E1$ contributions to the angle-integrated bremsstrahlung cross sections at a photon energy $E_\gamma=1$~MeV for the $\alpha+p$ system in the GCM~\cite{DE14} and NCSM/RGM approaches and for the $\alpha+n$ system in the GCM approach~\cite{DE14}.}
\end{minipage}
\hspace{1.7pc}
\begin{minipage}{18pc}
\includegraphics[width=18pc]{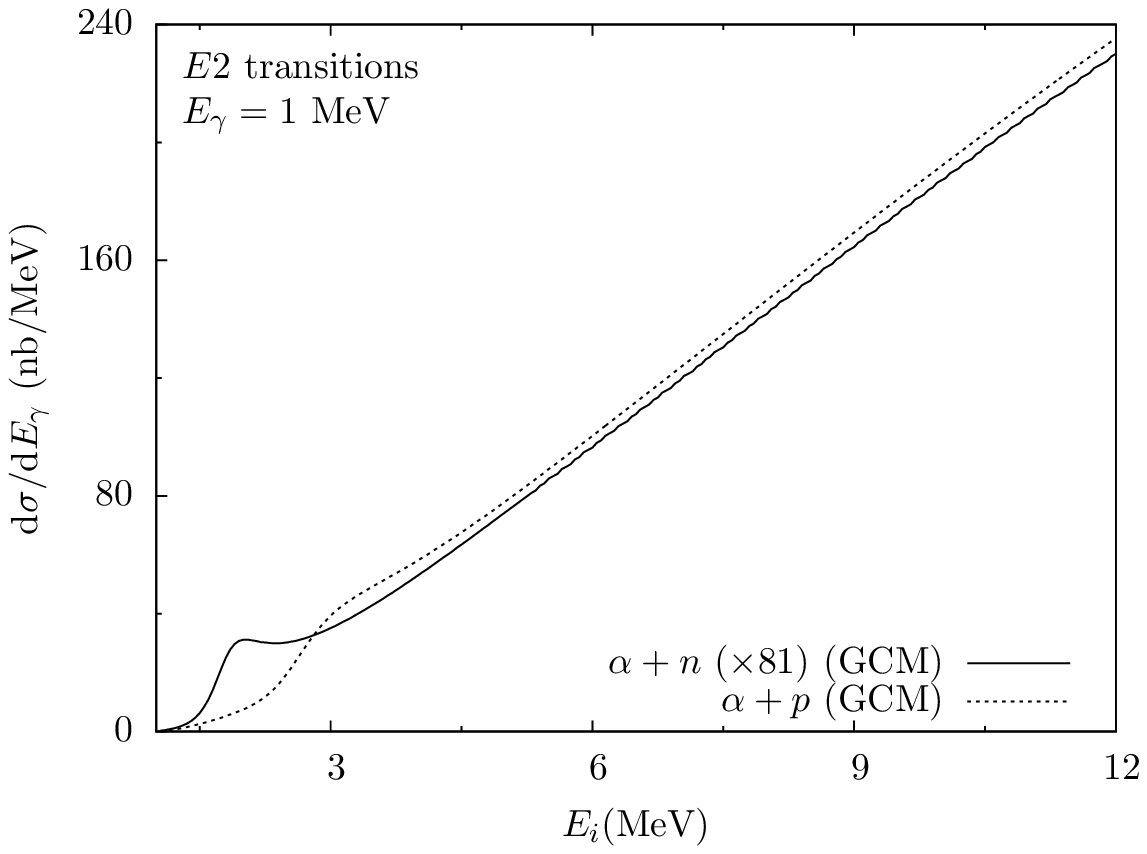}
\caption{\label{E2} The $E2$ contributions to the $\alpha+p$ and $\alpha+n$ angle-integrated bremsstrahlung cross sections at a photon energy $E_\gamma=1$~MeV in the GCM approach~\cite{DE14}. The $\alpha+n$ bremsstrahlung cross sections are multiplied by $81$.}
\end{minipage} 
\end{figure}

The $E1$ contributions to the angle-integrated bremsstrahlung cross sections at a photon energy $E_\gamma=1$~MeV for the $\alpha+p$ system in the GCM~\cite{DE14} and NCSM/RGM approaches and for the $\alpha+n$ system in the GCM approach~\cite{DE14} are displayed in Fig.~\ref{E1}.
Technical details about the GCM calculations can be found in \cite{DE14}.
The peaks in the bremsstrahlung cross sections are at energies which correspond to the final states at the $3/2^-$ resonance energies. 
The peak is at higher energy for the $\alpha+p$ system than for the $\alpha+n$ system since the $3/2^-$ resonance energy is higher for the $\alpha+p$ system than for the $\alpha+n$ system.
Off-resonance, the $\alpha+p$ and $\alpha+n$ bremsstrahlung cross sections are nearly the same.

The $E2$ contributions are calculated in the GCM approach for the $\alpha+N$ systems at the same
photon energy ($E_\gamma=1$~MeV) and are displayed in Fig.~\ref{E2}. For both systems, but especially for the
$\alpha+n$ system, the $E2$ transitions are much weaker than the $E1$ transitions. The ratio of the orders of
magnitude of the electric transition contributions between the $\alpha+p$ and $\alpha+n$ bremsstrahlungs is
roughly estimated by the square of the ratio of the effective charges of the $\alpha+p$ and $\alpha+n$ systems
which is 1 for the $E1$ transitions and 81 for the $E2$ transitions~\cite{DE14}.

For the $\alpha+p$ system, the $E1$ contributions to the angle-integrated bremsstrahlung cross sections,
at $E_\gamma=1$~MeV, are calculated in the NCSM/RGM approach, too. The maximum number of
quanta in the harmonic oscillator basis considered in this model is 13 and the oscillator frequency is
$20$~MeV/$\hbar$. The inter-nucleon potentials have been softened to minimize the influence of momenta
larger than $2.0$~fm$^{-1}$. Contrary to the study of the $\alpha+p$ elastic scattering performed in~\cite{HLN13}, only the cluster states including the ground state of the $\alpha$ particle are considered here. 
At $E_\gamma=1$~MeV, the $\alpha+p$ bremsstrahlung cross sections have the same order of magnitude in the GCM and NCSM/RGM approaches. 
The differences in the bremsstrahlung cross sections are probably due, for most part, to the differences in the $\alpha+p$ elastic phase shifts obtained with these approaches. 
Indeed, by considering only the ground state of the $\alpha$ particle in the NCSM/RGM basis, the $\alpha+N$ elastic resonances are not well reproduced by the NCSM/RGM. However, the agreement between the theoretical and
experimental elastic phase shifts can be improved by increasing the number of configurations in the
NCSM/RGM and/or including five-nucleon NCSM states in the description of the colliding wave functions,
like in the NCSM with continuum approach~\cite{BNQ13}. This work is in progress and should lead to
more precise bremsstrahlung cross sections.
\section*{Acknowledgments}
One of the authors (JDE) thanks D. Baye for useful discussions. This text presents research results
of the interuniversity attraction pole programme P7/12 initiated by the Belgian-state Federal Services
for Scientific, Technical and Cultural Affairs. A part of this work was done with the support of the
F.R.S.-FNRS and the support of LLNL under Contract DE-AC52-07NA27344. TRIUMF receives
funding via a contribution through the National Research Council Canada. This material is based in
part upon work supported by the U.S. Department of Energy, Office of Science, Office of Nuclear
Physics, under Work Proposal Number SCW1158.

\end{document}